\documentclass[usenatbib,referee]{amsart}

\usepackage{natbib}
\usepackage{color}
\usepackage{amsmath}
\usepackage{amsaddr}
\usepackage[figuresright]{rotating}
\usepackage{booktabs,caption,fixltx2e}
\usepackage[flushleft]{threeparttable}

\title[Covariance testing]{A robust covariance testing approach for high-throughput data}

\author{Yi-Hui Zhou}
\address{{$yihui\_zhou$@ncsu.edu}\\
Bioinformatics Research Center\\
Department of   Biological Sciences\\  North Carolina State University,
Raleigh, North Carolina, U.S.A.}

\begin{document}





\label{firstpage}

\begin{abstract}
{The problem of testing changes in covariance has received increasing attention in recent years, especially in the context of high-dimensional testing.  A number of approaches have been proposed, all limited to the two-sample problem and involving varying statistics and assumptions on the number of features $p$ vs. the sample size $n$. There are  no general approaches to test association of covariances with a continuous outcome.  We propose a uniform framework for testing association of covariances with an experimental variable, whether discrete or continuous. The approach is not limited by the data dimensions.  Our test procedure (i) does not rely on parametric assumptions; (ii) works well for a range of $p$ and $n$ (e.g., does not require $n > p$);  (iii) provides correct type I error control, and (iv) includes four different statistics, to ensure power and flexibility under various settings, including a new ``connectivity" statistic that is sensitive to changes in overall covariance magnitude. We demonstrate that, for the two-sample special case, the proposed statistics are permutationally equivalent or similar to existing proposed statistics. We demonstrate the power and utility of our approaches via simulation and analysis of real data.  The approach is implemented in an $R$ package. 
 }
\end{abstract}


\maketitle

\section{Introduction}

Tests of changes in covariance structure have long been available (\cite{john1971some}), and classical likelihood approaches require that the sample size $n$ to be large compared to the number of features $p$  (\cite{anderson1962introduction}). A number of investigators have recently re-considered the problem of testing equality of $p\times p$ covariance matrices $H_0: \Sigma_1=\Sigma_2$ based on samples of sizes $n_1$ and $n_2$, where $n_1+n_2=n$.  In settings where $p>min\{n_1,n_2\}$, likelihood ratio testing may perform poorly or be undefined.  Li and Chen
(\cite{li2012two}) derived an approximately standard normal statistic for the Frobenius norm of differences in the two $p\times p$ sample covariance matrices, with considerable attention to sources of bias when $p$ is large.
 \cite{cai2013two} proposed a maximum standardized difference statistic between two sample
covariances, with testing based on extreme value results.  
The two approaches are designed for very different alternatives, ranging from modest but widespread differences in the two sample covariance matrices (Li-Chen) to large differences in a very few covariance elements (Cai). \cite{peng2016more} proposed a method and theorem that apply to bandable covariance matrices.  These newer methods require only relatively weak assumptions on the original data, such as fourth-moment bounds (\cite{li2012two}) and tail constraints (\cite{cai2013two}) that are much weaker than normality. However, the asymptotic results are not always fully aligned with the goals - e.g., \cite{li2012two} derived asymptotics for increasing sample size, although the approach is intended for large $p$.  In addition,  small sample false-positive control has not been established for most of these methods.

The two-sample problem can be viewed as an ``association" of the covariance matrix with a binary
group indicator.  More generally, the investigator may be interested in trend association of covariance with an experimental variable $y$ that might be multi-level, or on a continuous scale. To our knowledge, no general method is available with the requisite flexibility, without parametric requirements or assumptions of the feature size $p$ relative to $n$. Moreover, existing methods have been published in isolation, providing little opportunity to contemplate power characteristics for various types of alternatives. 

 In this manuscript, we propose four different statistics, when it is anticipated that a change in $y$ will result in (1) a directional change in elements of the covariance matrix; (2) a non-directional change in covariance; (3) a change in the overall magnitude of covariances; (4) a large change in one or a few elements of the covariance matrix. In contrast to almost all of the relevant literature, the statistics apply naturally whether $y$ is continuous or discrete. Permutation is used to ensure control of type I error, and for statistics (1) and (3) the permutation moments are known, so that fast testing approximations are available.

This paper is organized as follows. In Section \ref{method}, we introduce the method and test statistics. Section \ref{theorem} establishes permutation equivalence between some of the proposed statistics and  existing methods for the two-sample problem. In Section \ref{simulation}, we compare the proposed  statistics with existing methods, in terms of type I error and power. Several different simulation settings are presented for the two-sample problem, comparing our statistics to existing methods.  In addition, we compare our proposed methods in the setting with continuous $y$. Finally, we apply the proposed methods to a methylation dataset, with findings that are biologically sensible. 

\section{Methods}\label{method}

Let $X$ be the $p \times n$ data matrix and $y$ be the $n$-vector of clinical/experimental data.
For a sample $x_{.j}$, we assume a zero mean vector and denote the covariance, which may depend on $y_j$, as $\Sigma_{y_j}$. The zero-mean assumption is implicit in most covariance tests, following an intent that the test statistics be sensitive only to changes in covariance.  In practice, residuals can be obtained for regression of each feature $x_{i.}$ on $y$, with the residuals used as a new data matrix.

To motivate our statistics, we adopt a conceptual trend model for the covariance dependence of $X$ on $y$: $\Sigma_y=\beta_0+\beta_1 y$ for $p\times p$ matrices $\beta_0$, $\beta_1$. Thus for sample $j$, 
according to our assumptions $E(X_{ij}|y_j)=0$ for each $i$, and
${\rm cov}(X_{ij} X_{i\prime j}|y_j)=E(X_{ij} X_{i\prime j}|y_j)= \beta_{0,i,i\prime}+\beta_{1,ii^\prime}y_j$.
Letting $z_{ii^\prime j}=x_{ij}x_{i\prime j}$, the model immediately suggests linear regression of $z$ on $y$, for which the least-squares slope solution is $\widehat{\beta}_{1,i i\prime}=(\Sigma_j z_j y_j/n-\overline{z}\overline{y})/(s^2_y \frac{n-1}{n})$.
We make two observations: (1) we do not consider $\beta_{0, i i^\prime}$ to be of interest for detecting covariance changes, and (2) linear rescaling of $y$ will not meaningfully change our results, because
it results in constant changes in the proposed statistics. Thus without loss of generality we  assume $\bar{y}=\sum_j y_j/n=0$, so $\widehat\beta_{1, i i^\prime}=\frac{1}{n s_y^2} \sum_j x_{ij} x_{i^\prime j} y_j=\frac{1}{n s_y^2} \sum_j z_{ii^\prime,j} y_j$.

These least squares solutions are not intended to be used directly (except for $M$ below), but serve to motivate global test statistics described below. 

To effectively measure the covariance changes, we  propose $S=\sum_i \sum_{i^\prime} \hat\beta_{1, i i^\prime}$ as a {\it summation} statistic to detect global changes in covariances that are  concordantly associated with the experimental variable $y$ (i.e., in the same direction).
In contrast,  $Q=\sum_i \sum_{i^\prime} \widehat\beta_{1, i i^\prime}^2$ is a {\it quadratic form} statistic
that is sensitive to changes that are not directionally concordant.
Useful simplifications for these two statistics are as follows. We have
\begin{eqnarray*}
S& =& \sum_i \sum_{i^\prime}
  \sum_j x_{ij} x_{i^\prime j} y_j= \sum_j y_j \sum_i x_{ij} \sum_{i^\prime} x_{i^\prime j}= \sum_j y_j (\sum_i x_{ij})^2 = \sum_j w_j y_j =y^T w 
\end{eqnarray*}
for $w_j=(\sum_i x_{ij})^2$.
Also,
\begin{eqnarray*} 
Q & = & \sum_i \sum_{i^\prime} (\sum_j x_{ij} x_{i^\prime j} y_j)^2 =  \sum_i \sum_{i^\prime} \sum_j \sum_{j^\prime} x_{ij} x_{i^\prime j}  x_{i j^\prime} x_{i^\prime j^\prime} y_j y_{j^\prime} \\
&=&\sum_j\sum_{j^\prime} y_j y_{j^\prime} \sum_i x_{ij} x_{i j^\prime} \sum_{i^\prime} x_{i^\prime j} x_{i^\prime j^\prime} =\sum_j\sum_{j^\prime} y_j y_{j^\prime} a_{j j^\prime}
\end{eqnarray*}
 where $a_{j j^\prime}=(\sum_{i} x_{ij} x_{i j^\prime})^2$.  We use the superscript ``${\circ k}$" to denote the element-wise exponent of a matrix to power $k$, and the matrix with elements $a_{j j^\prime}$ can be simplified to $A=(X^T X)^{\circ 2}$. Finally, we have $Q=y^T A y$
(a quadratic form).

For both $S$ and $Q$, the initial motivation based on $p\times p$ covariance terms results in statistics that ultimately use $n$-vectors and $n\times n$ matrices.  We note that approximate permutation $p$-values for $S$, an inner product, can be computed using the moment-correlated correlation (MCC) method of \cite{zhou2015hypothesis}, avoiding the computational cost of direct permutation.
For $Q$, the permutation quadratic form moment method of \cite{zhou2013space} does not apply, as the row/column sums of $A$ are not constant, and thus direct permutation must be used to obtain $p$-values.

Each element $a_{j j^\prime}$ has the form of a squared correlation between samples $j$, $j^\prime$, and so e.g. $b_j=\sum_{j^\prime} a_{j, j^\prime}$ reflects broad-scale correlation (a ``connectivity index") of sample $j$ with remaining samples.  We propose the {\it connectivity} statistic $C=y^T b$ to reflect correlation between $y$ and the connectivity index.  Correlations between samples are ultimately driven by correlation between features, and $C$ reflects the tendency for the aggregate magnitude of feature-feature correlations to be associated with $y$, which is quite different from the type of alternative envisioned for $S$ and $Q$. For $C$ we can also use MCC as an alternative to direct permutation for computing $p$-values.


Our fourth statistic was inspired by 
\cite{cai2013two}, who devised a test for the maximum element difference, scaled by an appropriate standard error, for sample covariance
matrices in the two-sample problem.  Again, we wish to generalize the statistic, 
and propose the {\it maximum} statistic $M$ =$max_{i, i^\prime} \frac{|\widehat\beta_{1,ii^\prime}|}{SE_{ii^\prime}}$, and $SE_{i,i^\prime}$ is the standard error for $\hat\beta_{1,ii^\prime}$.
Under the null, $\widehat\beta_{0,ii^\prime}\approx\bar {z}_{i i\prime}=\sum_j z_{i i\prime j}/n$, then the approximate residual is $\widehat{\epsilon_{i i\prime j}}=z_{i i\prime j}-(\widehat\beta_{0,i i\prime}+\widehat\beta_{1,i i\prime} y_j)$. Therefore an approximate standard error is $SE_{i i\prime}=\sqrt{\frac{\sum_j \widehat \epsilon^2_{i i\prime j}/(n-2)}{\sum_j (z_{i i\prime j-\bar{z_{i i\prime}}})^2}}$. The null distribution of $M$ is evaluated by permutation, which can be computationally
intensive for large $p$.

\subsection{Permutation testing}
Letting $\Pi$ denote a random permutation of $n$ elements from among the $n!$ possibilities (realized value $\pi$), the statistics for permutation $\pi$ are  
$S_{\pi}=y_\pi^T w$, $Q_\pi=y_\pi^T A y_\pi$, $C_\pi=y_\pi^T b$, and $M_\pi$ (which requires
computation of the $\hat\beta_1$ values and standard errors for each permutation).
$S$ and $C$ are subjected to two-sided testing, with $p$-values based on both right and left tails, while $Q$ and $M$ are one-tailed, rejecting for large values.  For example, with $K$ random permutations
and $\pi[k]$ denoting the $k$th permutation, the empirical $p$-value for $S$ is
$p_S=\sum_{k=1}^K I[|S_{\pi[k]}|\ge |S_{observed}|]/K$, while the $p$-value for $Q$ is
$p_Q=\sum_{k=1}^K I[Q_{\pi[k]}\ge Q_{observed}]/K$.

The null hypothesis is that the relationships of columns of $X$ to the elements of $y$ are 
exchangeable (\cite{good2002extensions}), which holds if $X$ and $y$ are drawn from independent
distributions. 
A primary advantage of permutation testing is that, aside from slight issues due to discreteness or tied outcomes, type I error rates are controlled without requiring parametric assumptions
(\cite{zhou2015hypothesis}). This property is especially important for covariance association testing, enabling implementation for data of any size $p$ and $n$.

\section{Two group comparisons and permutation equivalence}\label{theorem}

For a subset $\omega$ of $n_\omega$ samples, we consider the $p\times p$ noncentered sample covariance $\widehat{\Sigma}_\omega= X_\omega X_\omega^T/n_\omega$, under the justification that the population mean for each data row is assumed to be zero.  A single $i, i^\prime$ element is 
$\hat\sigma_{i i^\prime, \omega}=(\sum_{j\in\omega}x_{ij}x_{i^\prime j})/n_\omega$.
The following result ties our proposed statistics to two reasonable statistics in comparing two sample covariance matrices.

\medskip
\noindent \textbf{Result 1.} Let $\omega_1$ and $\omega_2$ be the indexes for samples in groups 1 and 2, respectively, and the subscripts 1 and 2 will be used for simplicity. We assign the experimental variable
$y_j = \frac{1}{n_1}$ if $j\in\omega_1$, and $y_j=\frac{-1}{n_2}$ if $j\in\omega_2$. We use $\xi$ to denote the operator that sums all elements of a matrix. 
\begin{itemize}
\item[(i)] The directional statistic $S$ is equivalent to $\xi(\widehat{\Sigma}_1-\widehat{\Sigma}_2)$,
\item [(ii)] The non-directional statistic $Q$ is equivalent to  $\xi \bigl((\widehat{\Sigma}_1-\widehat{\Sigma}_2)^{\circ 2}\bigr)$
\end{itemize}

\noindent
{\it Proof.} We have covariance element differences 
\[\hat\sigma_{i i^\prime,1}-\hat\sigma_{i i^\prime,2}=\frac{\sum_{j\in\omega_1} x_{ij}x_{i^\prime j}}{n_1}-\frac{\sum_{j\in\omega_2} x_{ij}x_{i^\prime j}}{n_2}=\sum_j x_{ij} x_{i^\prime j} y_j=\hat\beta_{1,i i^\prime}.
\]
Summing over the $p\times p$ elements we have $\xi(\widehat{\Sigma}_1-\widehat{\Sigma}_2)=\sum_i\sum_{i^\prime} \hat\beta_{1, i i^\prime}$, and $\xi\bigl((\widehat{\Sigma}_1-\widehat{\Sigma}_2)^{\circ 2}\bigr)=\sum_i\sum_{i^\prime} \hat\beta_{1, i i^\prime}^2$. 

We note that $\xi\bigl((\widehat{\Sigma}_1-\widehat{\Sigma}_2)^{\circ 2}\bigr)$ is essentially the Frobenius norm statistic proposed by \cite{li2012two}, except that the authors employed various bias corrections (because $E(\hat\Sigma)\ne \Sigma)$
 to construct their statistic.  When using permutation, such corrections are unnecessary, because the observed and permuted values are subject to the same bias.  Additionally, two different statistics
will provide permutation $p$-values that are identical if the statistics are {\it permutationally equivalent} as defined in section 2.4 of \cite{pesarin2010permutation}.  Thus it is immaterial that the Frobenius norm 
involves a square root not used in the statistic shown here (\cite{golub2012matrix} pg. 55). Moreover, there is no need for standard error estimation, e.g. as employed by \cite{li2012two} to compute an approximately $N(0,1)$ statistic.

\begin{figure}[t]
\begin{center}
\includegraphics[width=12cm]{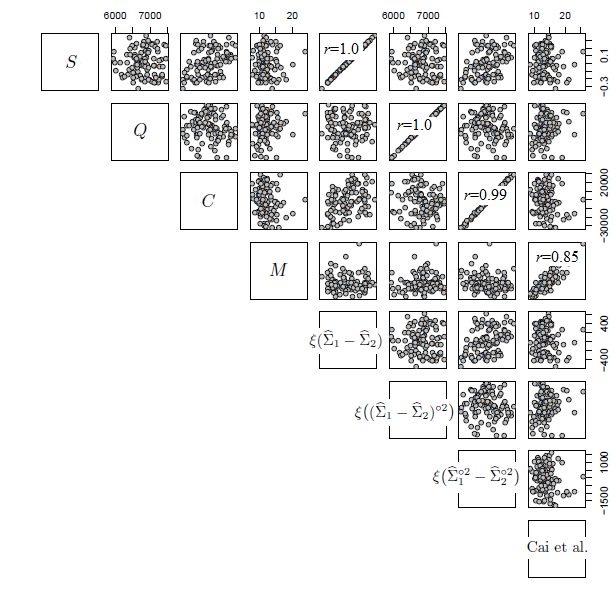}
\caption{Comparison of the four proposed statistics to various existing statistics for the two-sample problem, for a single simulated dataset and 100 permutations. Pearson correlations illustrate the exact and approximate correspondence of some pairs of statistics.}
\label{perm}
\end{center}
\end{figure}

For the two-sample problem, the number of unique outcomes is ${{n_1+n_2}\choose{n1}}<n!$, but the general approach of drawing from the $n!$ permutation possibilities is still valid. Figure \ref{perm} shows the results from 100 random permutations of $y$ for the two sample problem with $n_1=n_2=20$, $p=50$. A single $X$ was generated using the null version of Model 2 described in the next section, but the qualitative results hold regardless of the choice of $X$.
 As we showed in the results, $S$ and $Q$ are equivalent to $\xi(\widehat{\Sigma}_1-\widehat{\Sigma}_2)$ and $\xi\bigl((\widehat{\Sigma}_1-\widehat{\Sigma}_2)^{\circ 2}\bigr)$ respectively. Under the permutations, $C$ has high Pearson correlation over permutations with 
$\xi\bigl(\widehat{\Sigma}^{\circ 2}_1-\widehat{\Sigma}^{\circ 2}_2\bigr)=
\xi(\widehat{\Sigma}^{\circ 2}_1)-\xi(\widehat{\Sigma}^{\circ 2}_2)
$, supporting the perspective that $C$ reflects a contrast in the overall magnitude of covariances. Finally, our $M$ is correlated under permutation with the statistic from \cite{cai2013two}, although they differ modestly due to differences in the standard errors used.  

This permutation example underscores the correspondence between our statistics and those that seem ``natural" for the two-sample problem, but we re-emphasize that our statistics apply for either discrete or continuous $y$.

\section{Type I error and Power}\label{simulation}
We use permutation to obtain $p$-values for our proposed statistics, and therefore we expect accurate type I error control under the exchangeable null hypothesis.  However, ties and slight discreteness issues could potentially influence false positive rates. Competing methods rely on asymptotics to obtain $p$-values, and thus should be examined carefully.  For the two sample problem, we start by examining the type I error and power characteristics for the proposed and existing statistics, and follow with power analyses for the proposed statistics for some settings with continuous $y$.

\subsection{Two sample comparisons} \label{twosample}

Initial comparisons follow the simulation settings from \cite{li2012two}, for which feature covariances were described using auto-regressive notation. More compactly, we describe their simulation settings in terms of the covariance matrices.

\subsubsection{Simulation Model 1 (type I error)}

We assume the first population $X_1$ $\sim N(0, \Sigma_1)$; while the second population $X_2$ $\sim N(0, \Sigma_2)$, where 

$\Sigma_{1ij} = \begin{cases} 1+\theta^2_1, & \mbox{if } i\mbox{  $=j$} \\ \theta_1, & \mbox{if } i\mbox{ $= j+1  or  j-1$} \\
0, & \mbox{if } i \mbox { $ \neq j-1,j,j+1$} \end{cases}$,
$\Sigma_{2ij} = \begin{cases} 1+\theta^2_1+\theta^2_2, & \mbox{if } i\mbox{ $ =j$} \\ \theta_1(1+\theta_2), & \mbox{if } i\mbox{ =$ j+1 \ or  j-1$} \\
0, & \mbox{if } i\mbox { $ \neq j-1,j,j+1$}\end{cases}$.

\noindent The difference between the two covariance matrices is 
$$\Sigma_{0ij}-\Sigma_{1ij} = \begin{cases} \theta^2_2, & \mbox{if } i\mbox{ $=j$} \\ \theta_1 \theta_2, & \mbox{if } i\mbox{ $= j+1 \ or  j-1$} \\
0, & \mbox{if } i\mbox { $ \neq j-1,j,j+1$}\end{cases}.$$ To assess type I error, we set $\theta_2=0$, which implies the null $\Sigma_{1}-\Sigma_{2}=0$. We show results for
$n_1=n_2=\{20,50,80,100\}$ and feature dimension $p=\{32,64,128,256,512,700\}$. The number of simulations was 1000 for each setting, and 1000 permutations for the permutation methods.

\begin{figure}[t]
\begin{center}
\includegraphics[width=15cm]{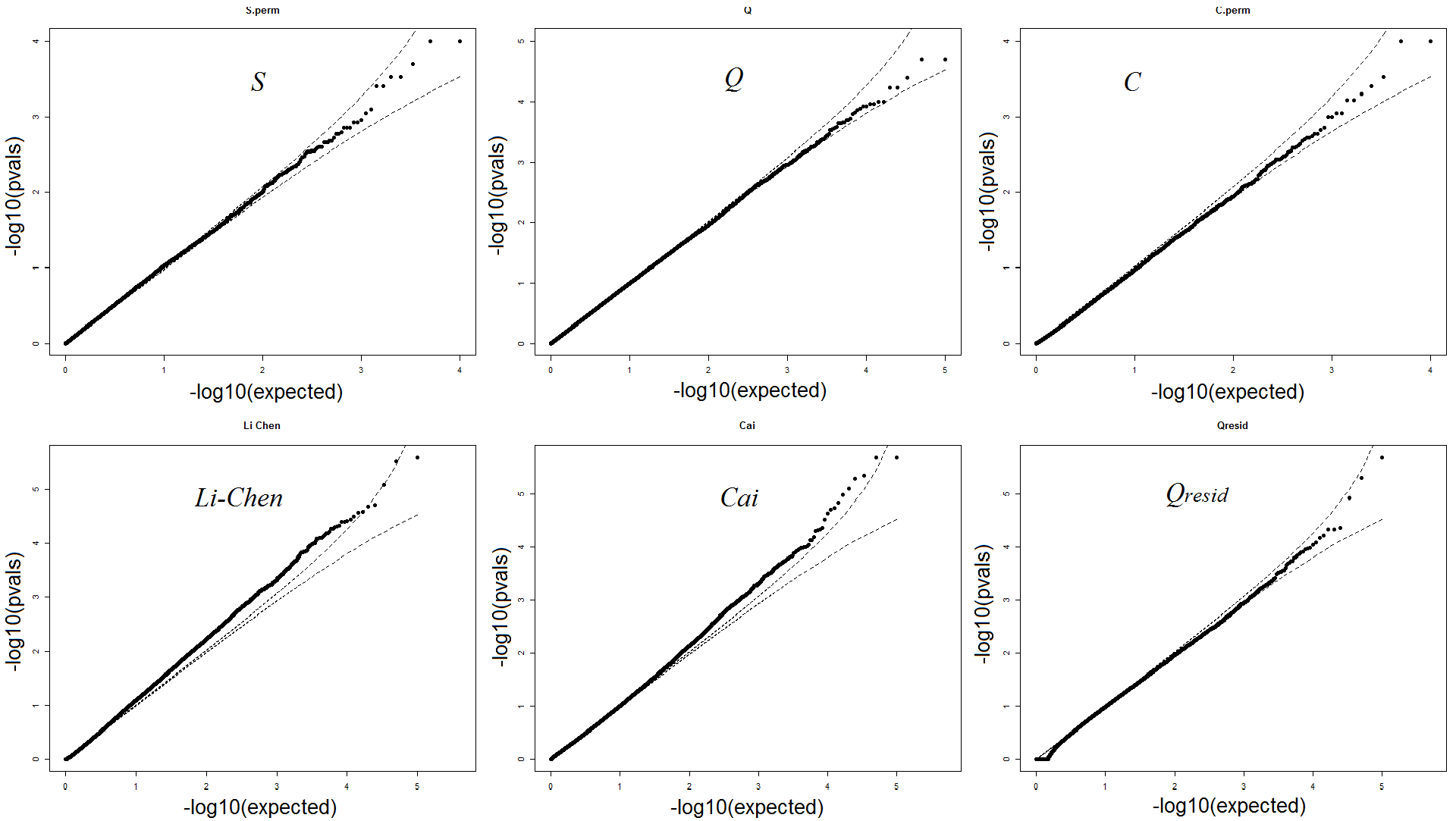}
\caption{QQplots for several of the proposed methods, as well as existing methods, for the null two-sample problem of Simulation Model 1, $p=50$, $n_1=n_2=20$.}
\label{Fig:qqplots}
\end{center}
\end{figure}

Table \ref{tab:typeInormal} shows that for this multivariate normal model, most methods perform well and control type I error.  The exchangeability hypothesis holds, and so $S$, $Q$, and $C$, based on permutation, would be expected to perform well.
The $Cai$ (\cite{cai2013two}) method is noticeably anticonservative for $\alpha=0.05$ for the smaller sample size ($n_1=n_2=20$), and more so as $p$ increases.  For the setting with $n_1=n_2=20$, $p=50$, 100,000 simulations were performed to provide greater insight into tail behavior (Figure \ref{Fig:qqplots}). $p$-values for the proposed permutation methods perform well.  The residualized $Q$ (lower right panel) shows that residuals from the linear regression of $X$ on $y$ can be used, provided the residualization is also performed inside the permutation loop.

\subsubsection{Simulation Model 2 (type I error)}

Here we follow the previous Simulation Model but with skewed data elements.  Specifically, let $G(w;4,0.5)$ denote the gamma distribution function with shape parameter 4 and scale 0.5 evaluated at $w$.  Then if $W\sim G$, $X=W-2$ has mean zero and variance 1, i.e. follows a centered gamma.  The elements of $X_1$ and $X_2$ are drawn as shown above, following the same null covariance structure that was used in Simulation Model 1.

Here the $Cai$ approach in \cite{cai2013two} becomes conservative, both with increasing sample size and  feature size (Table \ref{tab:typeIgamma}).  The $Li-Chen$ method is anti-conservative, but the type I error becomes closer to nominal as the sample size and feature size increase.
%
As expected, our proposed methods are very robust in controlling type I error for all $n$ and $p$, as the skewness in data elements does not violate the exchangeability property.

\subsubsection{Simulation Model 3 (power)}

For power comparisions, we return to the multivariate normal data elements. We use Simulation Model 1, but with covariance matrices determined by $\theta_1=2, \theta_2=1$ (one of the simulation models also used by \cite{li2012two} and summarized in their Table 4). Although this simulation model was used by \cite{li2012two} to support their proposed statistic, our proposed $C$ has consistently highest power for all the $n, p$ settings.  The $Li-Chen$ statistic shows power slightly higher than that of $Q$, even though
they both are based on the Frobenius norm. We speculate that the reason is related to the fact that permutation testing is conditional on the observed data, and the power difference nearly disappears at the larger sample sizes.  It is perhaps a bit surprising that $S$ is less powerful than $Q$, as the covariance differences are directional. However, the squared terms in $Q$ also may effectively act to reduce noise, and we later show situations in which $S$ is more powerful.  The $Cai$ and $M$ statistics show the lowest power, as they use only the most extreme covariance difference element, and do not aggregate over the large number of covariance difference elements.

\subsection{Simulations with a continuous $y$} \label{cont}

\subsubsection{Simulation Model 4.}
\begin{figure}[t]
\begin{center}
\includegraphics[width=15cm]{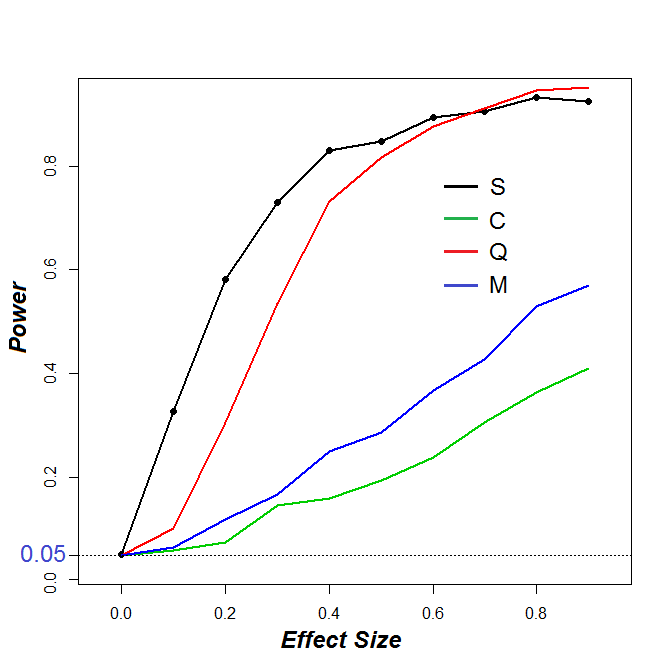}
\caption{Power comparision among $S$, $Q$, $C$, $M$ for Simulation Model 4. The dashed line at $\alpha=0.05$ indicates that all the proposed methods control type I error well under the null ($\rho=0$). The effect size $\rho$ ranges from $0$ to $0.8$. }
\label{Fig:power_conty}
\end{center}
\end{figure}

For this simulation model, values in $y$ are drawn iid $N(0,1)$ in each simulation, and converted to
the re-scaled experimental variable $y^*=\frac{y-min(y)}{max(y)-min(y)}  \in [0,1]$.
$X$ is drawn as multivariate $N(0,\Sigma_{y^*})$, with
$\Sigma_{y^*}=(1-y^*) \gamma_1 + \gamma_2$.  We assume $\gamma_1$ is the identity matrix and $\gamma_2$ is the compound symmetric matrix, 

$\gamma_{2ij}=\begin{cases} 1, \ if \  i=j \\ \rho,\ if \  i= j+1  \ or\  j-1 \\
0,\ if \  i  \neq j-1,j,j+1  \end{cases}$, 

\noindent in which we call $\rho$ the `Effect Size'.
 Under the null, there is no change in the covariance structure,  i.e.  $\gamma_2$ is the identity matrix, as is $\Sigma_{y^*}$ for all $y^*$. As $\rho$ increases, the relationship between the covariance and $y^*$ becomes stronger.
Figure \ref{Fig:power_conty} shows that the power for the proposed statistics is near the intended
 $\alpha=0.05$ when $\rho=0$. Figure \ref{Fig:power_conty} also shows that the directional statistic $S$ is the most powerful approach overall. 

\subsubsection{Simulation Model 5}
\begin{figure}[t]
\begin{center}
\includegraphics[width=15cm]{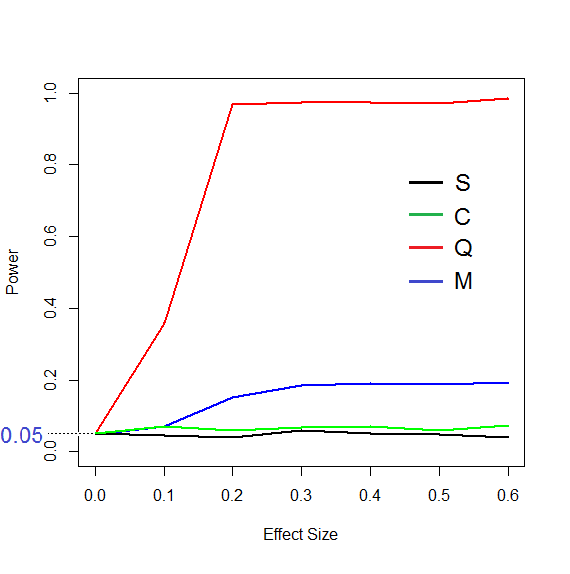}
\caption{Power comparision among $S$, $Q$, $C$, $M$ for Simulation Model 5. The dashed line at $\alpha=0.05$ indicates that all the proposed methods control type I error well under the null ($\rho=0$). The effect size $\rho$ ranges from $0$ to $0.6$. $Q$ is the most powerful method among these four for this simulation model.}
\label{Fig:power_conty2}
\end{center}
\end{figure}

This simulation model is a bit more complex, following a similar approach used in \cite{cai2013two}.
The approach generates covariance matrices that are non-directional in relationship to $y$.
 and with no overall variation in magnitude, while respecting the need for positive definiteness. 
  To an initial $p\times p$ identity matrix $I$, $\Sigma^{*(1)}$ was formed by drawing the first $p/2 \times p/2$ off-diagonal elements from $U[-\rho,\rho]$, followed by $\Sigma^{*(2)}=\Sigma^{*(1)}+\Sigma^{{*(1)}T}$, and $\Sigma_1=\Sigma^{*(2)}+(\lambda_{min}(\Sigma^{*(2)})+0.05) I$. $\Sigma_2$ is formed by reversing the rows and columns of $\Sigma_1$, and finally $\Sigma_{y^*}=\Sigma_1 (1-y^*)+\Sigma_2 y^*$, where $y^*$ is the result of  linear rescaling of $y$ to the $[0,1]$ interval as in the previous subsection.
Here $\rho\in[0,1)$ serves as an effect size, and $\Sigma_1$ and $\Sigma_2$ differ in the groups of genes that show correlation structure, but otherwise are the same in the average magnitude of elements and show no directionality. Figure \ref{Fig:power_conty2} provides the power comparision among the four proposed methods. As expected, $S$ and $C$ have little or no power, while $M$ has extremely modest power.
The statistic $Q$ benefits from aggregation of covariance squared differences, and thus has much more power than the other methods.  All methods control type I error properly (dashed line at 0.05 in Figure \ref{Fig:power_conty2}).

\section{Analysis of a methylation dataset}
We illustrate the methods via the analysis of the methylation dataset published by \cite{tsaprouni2014cigarette}. The data (GSE50660, Illumina Infinium HumanMethylation 450 BeadChip) consist of peripheral blood methylation signal from each of 464 individuals, with the binary experimental variable contrasting $179$ who never smoked ($y=0$) compared to $285$ former/current smokers ($y=1$).  The BeadStudio quantitative methylation signal was averaged for methylation sites within each of 22,003 genes using site-to-gene annotation, followed by annotation of genes to each of 4,512 Gene Ontology Biological Process terms, using Bioconductor v. 3.3. The methylation signal for genes within each BP term were used as data matrices for each analysis, as a ``pathway" analysis for changes in covariance signal to be associated with smoking status. The data matrices were first residualized using linear regression of each row $x_{i.}$ on $y$, in order to identify BP terms that were specifically associated with changes in covariance rather than changes in means.  Analysis using $S$, $Q$, and $C$ proceeded with 10,000 permutations for each BP term, and with residualization also performed inside the permutation loop.  

\begin{figure}[t]
\begin{center}
\includegraphics[width=15cm]{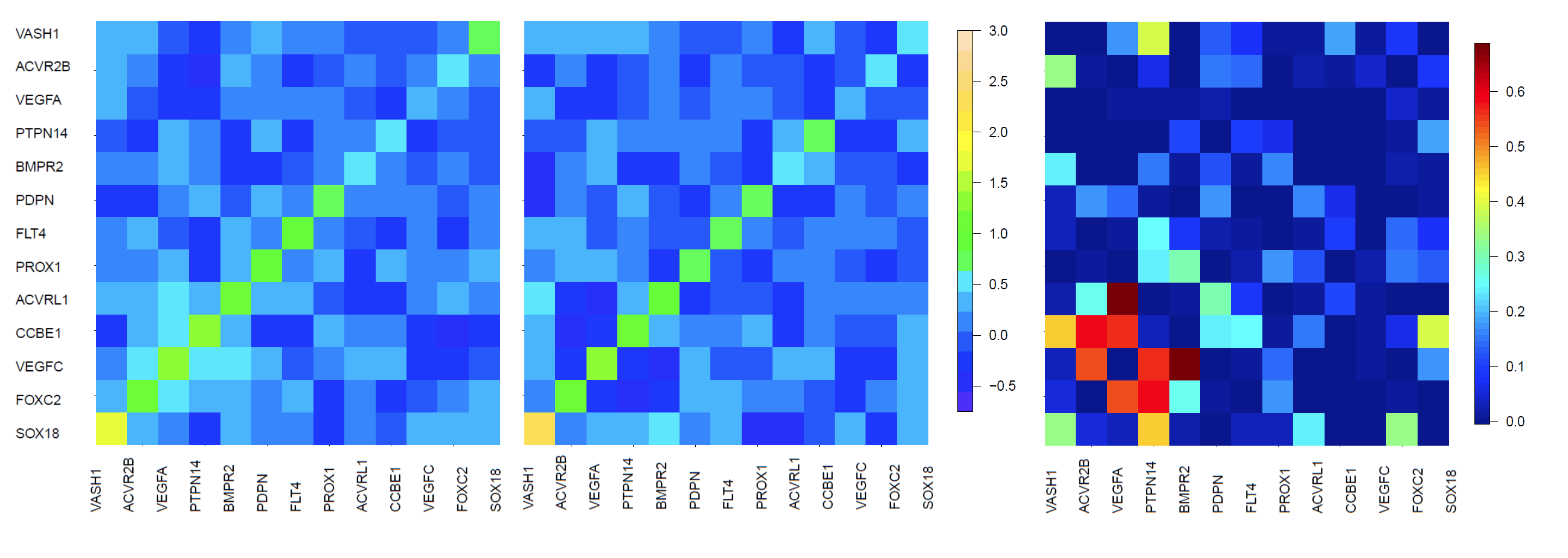}
\caption{Heatmap plots of the covariance matrices. The left panel is the covariance matrix for population 1; the middle plot is for population 2; the right one is based on the population difference.}
\label{heatmap}
\end{center}
\end{figure}

Among the Biological Process terms, two stood out as exceptional, showing no permutations more extreme than the original data: GO:0001946 (13  genes, lymphangiogenesis) and GO:0036303 (13 genes, lymph vessel morphogenesis), resulting in multiple comparison Benjamini-Hochberg $q<0.05$.  Sample covariance matrices for non-smokers and current/former smokers are shown in Figure \ref{heatmap}, along with the squared difference matrix.  For these same BP terms, the $C$ statistics were also modestly significant, with $p=0.0007$ for both terms, and with the current/former smokers showing an overall higher magnitude of covariance than the never smokers.  The results offer intriguing evidence of possible dysregulation of methylation -- the sensitivity of lymphangiogenesis to smoking has been recently described in urothelial cancer \cite{miyata2015smoking} and as part of chronic obstructive pulmonary disease pathogenesis \cite{hardavella2012lymphangiogenesis}. In addition, methylation of {\it NPR2}, a gene serving as a vascular endothelial
growth factor, has been highlighted as associated with maternal smoking in a recent large meta-analysis (\cite{joubert2016dna}).

\section{Discussion}
We have proposed four covariance test statistics, in a straightforward trend-testing framework that applies to general $y$. 
The approach is not limited by $p$, $n$, or whether $y$ is discrete or continuous. The availability of a testing for continuous $y$ is a distinct advantage over previous methods, making covariance testing a simple approach that can be applied in a huge variety of settings.  

The software is available upon request from the author.

\section{Acknowledgments}
This work is supported by R21HG007840 and EPA STAR RD83574701.

\bibliography{yihui}
\clearpage

\clearpage

\begin{table}[h!]
  \centering
  \caption{Type I error comparison, Sim. Model 1, $X_k \sim N(0,\Sigma_k)$, $\Sigma_1=\Sigma_2$}
  \label{tab:typeInormal}
  \begin{tabular}{c|c|cccccc}
\hline
\hline
    $n_1=n_2$ & Method &  p=32 & p=64 & p=128  & p=256  & p=512  & p=700\\
\hline
20&$S$ &0.053&0.041&0.047& 0.059& 0.065& 0.042\\
& $Q$ &0.055&0.058&0.046& 0.043& 0.043& 0.052\\
&$C$ &0.052&0.053&0.053& 0.059& 0.045& 0.058\\
& $Li-Chen$ &0.044&0.054&0.051& 0.048& 0.051& 0.038\\
& $Cai$ & 0.092&0.14&0.139&0.204&0.211&0.263\\
& $M$ & 0.053& 0.054 & 0.050 & 0.052 &0.051 &0.050\\
\hline
50& $S$ &0.056 &0.052& 0.051& 0.054& 0.053& 0.040\\
&$Q$ &0.052 &0.041& 0.045& 0.049& 0.042& 0.046\\
&$C$ &0.044 &0.049& 0.052& 0.037& 0.046& 0.054\\
&$Li-Chen$ &0.052 &0.060& 0.033& 0.043& 0.054& 0.049\\
&$Cai$&0.042&0.068&	0.058&0.065&0.055&0.059\\
& $M$ & 0.059 & 0.054& 0.051 & 0.048 & 0.051&0.050\\
\hline
80 & $S$ & 0.041 &0.055& 0.049& 0.052& 0.043& 0.050\\
& $Q$ & 0.065& 0.051& 0.040& 0.046& 0.044 &0.048\\
& $C$ & 0.041 &0.052& 0.058 &0.050& 0.043 &0.047\\
& $Li-Chen$ & 0.054 &0.060& 0.047& 0.048& 0.052& 0.053\\
& $Cai$ & 0.052	&0.056&	0.043&	0.052&	0.058&	0.041\\
& $M$ & 0.046 & 0.046 & 0.051 &0.047 & 0.050&0.049\\
\hline
100 & $S$ & 0.057& 0.052& 0.057& 0.058& 0.049& 0.059\\
& $Q$ & 0.039& 0.051& 0.050& 0.040& 0.060& 0.053\\
& $C$ & 0.035& 0.041&0.049& 0.047& 0.048& 0.054\\
& $Li-Chen$ &0.056& 0.049& 0.052& 0.046& 0.049& 0.048\\
& $Cai$ &0.05&0.052&0.043&0.039&0.036&0.047 \\
& $M$ & 0.047 & 0.054 & 0.048 & 0.050&0.048& 0.046\\
\hline
\hline 
  \end{tabular}
\end{table}

\begin{table}[h!]
  \centering
  \caption{Type I error comparison, Sim. Model 2, $\Sigma_1=\Sigma_2$, elements following centered gamma}
  \label{tab:typeIgamma}
  \begin{tabular}{c|c|cccccc}
\hline
\hline
    $n_1=n_2$ & Method &  p=32 & p=64 & p=128  & p=256  & p=512  & p=700\\
\hline
20 & $S$ & 0.068& 0.051& 0.042& 0.042& 0.048& 0.043\\
& $Q$ &0.057 &0.046& 0.066& 0.050& 0.047& 0.042\\
& $C$ & 0.040 &0.055& 0.045& 0.039& 0.067& 0.051 \\
& $Li-Chen$ & 0.158 &0.112&0.083&0.071&0.053&0.063\\
& $Cai$& 0.048&0.048	&0.058	&0.055	&0.083	&0.085\\
& $M$ &0.039&0.039&0.050&0.037&0.042&0.046\\
\hline
50& $S$&0.051 &0.059& 0.051 &0.055& 0.049& 0.047\\
&$Q$&0.055& 0.049& 0.055& 0.043& 0.051 &0.042\\
&$C$&0.038 &0.051 &0.038 &0.046& 0.052 &0.047\\
&$Li-Chen$&0.048	&0.048	&0.058&	0.055	&0.083	&0.085\\
& $Cai$ &0.016	&0.013&	0.010&	0.007	&0.003&	0.004\\
& $M$ &0.051 & 0.05& 0.055 & 0.042&0.050 &0.048\\
\hline
80& $S$ &0.051 &0.043 &0.054 &0.053 &0.052 &0.044\\
& $Q$ &0.056 &0.048 &0.043 &0.042 &0.038 &0.057\\
& $C$ & 0.057 &0.045 &0.041 &0.045 &0.057 &0.052\\
& $Li-Chen$ & 0.165&0.141&0.090&0.059&0.051&0.056\\
&$Cai$ & 0.019&0.010&	0.005&	0.006&	0.005&	0.002\\
& $M$ & 0.059 & 0.039 & 0.045& 0.045& 0.048&0.051 \\
\hline
100& $S$ & 0.044 &0.055 &0.057 &0.051 &0.044& 0.038\\
& $Q$ & 0.045 &0.042& 0.049& 0.049& 0.051& 0.039\\
& $C$ & 0.048 &0.034& 0.046 &0.042 &0.040& 0.042\\
& $Li-Chen$ & 0.176&0.133&0.088&0.069	&0.050&	0.046\\
& $Cai$ &0.013&0.009&	0.007	&0.003	&0.003&	0.003\\
& $M$ & 0.059 & 0.041& 0.058 & 0.052&0.053&0.050\\
\hline
\hline 
  \end{tabular}
\end{table}

\begin{table}[h!]
  \centering
  \caption{Power comparison, Sim. Model 3, $X_k \sim N(0,\Sigma_k)$, $\Sigma_1\ne\Sigma_2$}
  \label{tab:powernormal}
  \begin{tabular}{c|c|cccccc}
\hline
\hline
    $n_1=n_2$ & Method &  p=32 & p=64 & p=128  & p=256  & p=512  & p=700\\
\hline
20 & $S$ & 0.184 & 0.184& 0.179 &0.194& 0.193& 0.204\\
&$Q$ &0.211 &0.231 &0.235& 0.234 &0.221 &0.213\\
&$C$ &0.629 &0.831 &0.963& 0.998 &   1.000    &1.000\\
&$Li-Chen$ &0.273	&0.273	&0.252	&0.285	&0.269	&0.272\\
& $Cai$ & 0.138	&0.140	&0.164	&0.204	&0.233	&0.282\\
& $M$ & 0.129 & 0.072 & 0.050& 0.061 & 0.083 & 0.054\\
\hline
50 & $S$ &  0.438 &0.462& 0.482& 0.504& 0.465& 0.489\\
& $Q$ & 0.705 &0.751& 0.803 &0.809& 0.772& 0.789\\
& $C$ & 0.983 &1.000 &1.000 &1.000& 1.000& 1.000\\
& $Li-Chen$ & 0.752	&0.800&	0.824&0.861&	0.839&	0.857\\
& $Cai$ & 0.234&0.163&0.146&0.136&0.104&0.084\\
& $M$ & 0.270 & 0.133& 0.092 & 0.122 & 0.034& 0.051\\
\hline 
80 & $ S$ & 0.664 &0.673& 0.703& 0.690& 0.675& 0.710\\
&$Q$&0.955 &0.972 &0.991 &0.995 &0.992 &0.992\\
&$C$&0.999 &1.000 &1.000 &1.000   & 1.000    &1.000\\
& $Li-Chen$ &0.941&0.980&0.992	&0.994	&0.996	&0.998\\
& $Cai$ &0.496	&0.420&	0.377	&0.316&	0.246	&0.189\\
& $M$ & 0.574 & 0.394 & 0.333& 0.242& 0.253 & 0.201\\
\hline
100& $S$& 0.780 &0.781 &0.771 &0.785 &0.785 &0.795\\
& $Q$ & 0.991 &0.997 &0.999& 1.000& 1.000& 1.000\\
& $C$ &1.000 &1.000 &1.000 &1.000 &   1.000   & 1.000\\
& $Li-Chen$ &0.997	&1.000&	0.999&	1.000&	1.000&	1.000\\
& $Cai$ &0.700	&0.652	&0.557&	0.508&	0.423	&0.406\\
& $M$ & 0.700 & 0.649 & 0.601 & 0.487 &0.375 & 0.374\\
\hline
\hline
  \end{tabular}
\end{table}
\end{document}